\documentstyle[prd,aps,eqsecnum,preprint]{revtex}
\begin{document}

\title{
EQUATION OF STATE FOR THE 2+1 DIMENSIONAL\\
GROSS-NEVEU MODEL AT ORDER 1/N}
 
\author{M. Modugno, G. Pettini}
\address{Dipartimento di Fisica, Universit\`a di Firenze,
I-50125 Firenze, Italy\\
and Istituto Nazionale di Fisica Nucleare, Sezione di Firenze,
I-50125 Firenze, Italy}
\author{R. Gatto}
\address{D\'epartement de Physique Th\'eorique, Universit\'e de Gen\`eve,
CH-121 Gen\`eve 4, Switzerland}

\maketitle
\vspace{0.5cm}
\centerline{UGVA-DPT 1997/07/984}
\vspace{0.5cm}

\begin{abstract}
We calculate the equation of state
of the Gross-Neveu model in 2+1 dimensions at
order $1/N$, where $N$ is the number of fermion species. 
We make use of a general formula valid for four-fermion
theories, previously applied to the
model in 1+1 dimensions. 
We consider both the discrete and continuous symmetry
versions of the model. 
We show that the pion-like
excitations give the dominant contribution at low
temperatures. The range of validity for such pion
dominance is analyzed. The complete analysis from low to high
temperatures also shows that in the critical region
the role of composite states is relevant, even for quite large
$N$, and that the free-component behaviour at high $T$ starts
at about twice the mean field critical temperature. 
\end{abstract}

\pacs{PACS number(s): 11.10.Wx, 11.30.Rd, 11.15.Pg, 12.38.Mh}
\narrowtext

\section{Introduction}

To quantify the contributions of fundamental and composite 
degrees of freedom to the thermodynamics of strongly interacting
matter is a very difficult task. 

On general grounds, one expects
that at very low temperatures the light and relatively weakly 
interacting pions
should dominate the partition function, whereas at high temperatures
the relevant degrees of freedom are believed to be
almost free quarks and gluons. 

Unfortunately, when studying chiral symmetry 
breaking and restoration in QCD, where
the non perturbative regime plays a major role, we are not able, at present, 
to describe in a quantitative way, starting from the fundamental
lagrangian, what happens in going from low to high temperatures.

Quantitative results concerning the low-temperature behaviour 
of the fermion condensate and of the pion decay constant for
massless quarks have been presented by Gasser and 
Leutwyler \cite{leutw}, who 
make use of an effective lagrangian for pions, assumed to dominate
the thermodynamics in this regime. 

Other studies, complementary to Ref.\ \cite{leutw}
and to lattice results, have been 
generally performed by means of models which encompass
some properties of QCD, with the nice feature 
of starting directly from the fundamental degrees of freedom. 
Nevertheless, they generally do not go beyond
a mean field description, neglecting the role of bosonic
fluctuations. 
This approximation is clearly unsatisfactory 
for low temperatures, and also it only gives a very 
rough description of the transition. It is important to go 
beyond, at least to understand 
what ``low" and ``high" temperature mean, i. e. the regions 
where in QCD  only pions or only quarks and gluons 
would respectively dominate, and also to give an improved description 
of the region in between. 

Recently, a certain number of studies
have been devoted to this program, by means
of large-$N$ expansions \cite{coleman}, 
$N$ being the number of fermion
species, carried out directly at the level of the functional integral
describing the partition function. This procedure allows to preserve 
the desired symmetries from the beginning 
\cite{nzero}. 

At finite temperature, this technique
has been applied to the Nambu-Jona-Lasinio model \cite{kleva,nbeta}
and to the 1+1 dimensional Gross-Neveu model \cite{unosuenne,unosuenne1}. 
In spite of the approximations involved (and mainly the fact that
none of them starts from the fundamental QCD lagrangian),
these studies are useful 
towards a better description of chiral symmetry breaking and restoration. 

Here we present a calculation in the Gross-Neveu model in 2+1
dimensions. The model exhibits chiral symmetry breaking for
couplings stronger than a critical value, and it is renormalizable
in the $1/N$ expansion \cite{gross,rosen0,rosen1,parisi,ren}. 
The finite temperature extension of the model,
in the mean field approximation, has been
studied for instance in Ref.\ \cite{gn3mean,rosen}. 
At order $1/N$, the zero temperature effective potential has
been calculated in Ref.\ \cite{rosen2}, whereas
the finite temperature gap equation has been discussed, by using
the Ward-Takahashi identities, in Ref.\ \cite{shen}. 

We have calculated the finite temperature propagator of composite states
which enters a general formula valid for four-fermion
theories \cite{unosuenne} to calculate the finite 
temperature equation of state. 
This procedure had already been followed in studying the Gross-Neveu model
in 1+1 dimensions \cite {unosuenne1}. 
Here we perform, for Gross-Neveu in 2+1 dimensions, a full study of the equation
of state and we quantify the role
of the various contributions from low to high temperatures. 

We have studied both the version of the model with discrete chiral symmetry,
which is spontaneously broken by mass generation, and the version with
continuous chiral symmetry where, due to the
effective 2-dimensionality at finite temperature, there is no true phase 
transition. 
Thus we have included a bare mass which avoids I. R. divergencies in
this case. 

It is worth noticing that the full  
pressure at order $1/N$ is real for any temperature. 
This apparently trivial feature is not so obvious, since in general
the zero temperature effective potential is not always real for small values
of the order parameter, thus indicating either some instability
or ill-definiteness of the theory at this level 
\cite{unosuenne,rosen2,weinberg}. The fact that the value
of the effective potential at the absolute minimum turns out to be real
for any temperature allows for a safe  $N$-expansion
of the model.

In conclusion, by exploring the $1/N$ approximation
to the equation of state in the whole range of temperatures it
comes out that the naive expectation that pions
dominate the partition function at low temperatures is respected, 
as well as the
free-fermion dominance far above the critical temperature. 
Nevertheless the results show that the pion dominance is limited 
for temperature ranging from zero to about $0.1~T_c$, for a reasonable 
value of the chiral symmetry breaking term. 

Furthermore, in the critical region, where nonperturbative effects
are crucial, the mean field pressure is strongly modified 
by the $1/N$ correction, even for large $N$.

The rest of the paper is organized as follows: 
in Sect.\ \ref{sec:effpot} we recall the results for the renormalized
zero temperature effective potential, at order $1/N$.
In Sect.\ \ref{sec:press} we give the expression for the pressure
at order $1/N$ according to Ref. \cite{unosuenne} and briefly recall 
the solution of the mean field gap equation. 
In Sect.\ \ref{sec:results} we comment the results concerning the discrete
symmetry (Subsect. A) and the continuous symmetry (Subsect. B) versions of
the model.
Concluding remarks are in Sect.\ \ref{sec:concl}, whereas in 
Appendix\ \ref{sec:propag} we give the expressions for
the inverse bosonic propagators entering in the total pressure. 

\section{Effective potential at order ${\bf 1/N}$}
\label{sec:effpot}

In this paper we consider the four fermion model, 
in $D=2+1$ dimensions, defined by the following Lagrangian 
\cite{njl,gn}
\begin{equation}
{\cal L} = \bar{\psi}(i\gamma^{\mu}\partial_{\mu} - M)\psi 
+ {\lambda\over 2N} (\bar{\psi}\psi)^2 
- {\lambda_5\over 2N} (\bar{\psi}\gamma^5\psi)^2
\label{lagrangian}
\end{equation}
where $\psi$ is a multiplet of $N$ four-components fermion fields, 
and $\gamma^5$ is a traceless matrix which anticommutes with all 
$\gamma^{\mu}$ ($\mu = 0, 1, 2$) \cite{appel}. 
Notice that in dimensions $D=2+1$, in order to allow for chiral 
symmetry, we cannot consider the two-components spinorial 
representation of the Lorentz group $SO(2,1)$, since 
there is no $2\times 2$ matrix anticommuting
with the $\gamma^{\mu}$. On the contrary, in the case of 
four-components fermion fields, there are two matrices, $\gamma^3$ 
and $\gamma^5$,
which anticommute with $\gamma^0$, $\gamma^1$ and $\gamma^2$
(see {\sl e. g. } Ref.\ \cite{appel} for details). 

In the Lagrangian (\ref{lagrangian}) we have introduced two 
 couplings, $\lambda$ and $\lambda_5$, for the scalar and 
pseudoscalar part of the interaction respectively. 
In the following we consider the cases $\lambda_5=\lambda$ and 
 $\lambda_5=0$. Since the expressions for the effective potential
and for the pressure for $\lambda_5=\lambda$ can be easily reduced
to those of the $\lambda_5=0$ case by simply omitting the pseudoscalar 
contributions, in this section we derive 
all the formulas in the former case. 

For vanishing bare fermion masses, $M=0$, the Lagrangian (\ref{lagrangian})
is invariant under the discrete chiral transformation 
\begin{equation}
\psi \rightarrow \gamma^5 \psi
\label{discsim}
\end{equation}
for $\lambda_5=0$, 
and under the continuous chiral transformation
\begin{equation}
\psi \rightarrow {\rm e}^{\displaystyle i\theta\gamma^5}\psi
\label{contsim}
\end{equation}
for $\lambda_5=\lambda$. 
In the following, when considering the finite temperature extension
of the model with $\lambda_5=\lambda$, the mass term $M$ will be always
retained finite, in order to avoid I. R. divergencies, which would 
otherwise occur, according to well known no-go theorems \cite{coleman,rosen}. 
Furthermore, since we are interested in looking for hints for QCD,
where the quarks are massive, the exclusion of the case $M=0$ is not 
a severe limitation.

The expression for the effective potential at order $1/N$ is
\cite{unosuenne,rosen0,ren} ($\alpha\equiv M/\lambda$)
\begin{eqnarray}
{\cal V}^{n. r. }(\varphi) &=& {1\over2\lambda}\varphi^2 
+ \alpha\varphi - 2\int{dp^2\over (2\pi)^2}\int{d\omega\over 2\pi}
\ln\left(\varphi^2 +\omega^2 + {\bf p}^2 \right)\nonumber\\
&&+
{1\over N}\sum_{j=\sigma,\pi}
{1\over2}\int{d^2p\over (2\pi)^2}
\int{d\omega\over 2\pi}
~\ln\left[i{\cal D}_{0,j}^{-1}
(i\omega,{\bf p};\varphi)\right]
\label{poteff_nr}
\end{eqnarray}
where $i{\cal D}_{0,j}^{-1}$ is the zero temperature inverse 
bosonic propagator given in Appendix\ \ref{sec:propag}. 
The previous expression is renormalizable, since four fermion theories 
are renormalizable in the $1/N$ expansion \cite{gross,rosen0,rosen1,parisi,ren}. 
At the leading
order the U. V. divergent terms are proportional to $\varphi^2$,
while at the next-to-leading order there are divergencies proportional
to $\varphi^2$ and $|\varphi|^3$. To renormalize the expression 
(\ref{poteff_nr}) we introduce a three-momenta U. V. 
cutoff $\Lambda$ and write
\begin{equation}
{\cal V}(\varphi) = {\cal V}^{n. r. }(\varphi) + 
{1\over2\lambda}\varphi^2 \delta Z_2(\Lambda) 
+ |\varphi|^3\delta Z_3(\Lambda)
\end{equation}
To fix the finite part of the counterterms we follow the prescription
of Ref.\ \cite{ren}, from which one gets
\begin{equation}
{\cal V}(\varphi) = {\cal V}_0(\varphi) + 
{1\over N}\sum_{j=\sigma,\pi}{\cal V}_1^j(\varphi)
\label{zeroeffgen}
\end{equation}

\begin{equation}
{\cal V}_0(\varphi)= {1\over3\pi}|\varphi|^3 - {1\over2\pi}m_0 \varphi^2 
+ \alpha\varphi 
\label{voo}
\end{equation}

\begin{eqnarray}
{\cal V}_1^{\sigma}(\varphi) &=& {1\over 4\pi^2}
\int_0^{\Lambda}\!\! dy ~y^2\ln\left[{
|\varphi| - m_0 + 
\displaystyle{{1\over2}{ y^2 + 4\varphi^2\over y}}{\rm tan}^{-1}
\left(\displaystyle{ y\over 2|\varphi|}\right)
\over
\left|\pi y/4 - m_0\right|}
\right]\nonumber\\
&&\qquad\qquad
+{1\over 4\pi^2}
\left[ - 4\left(\Lambda +{4\over\pi}m_0\ln\Lambda\right)\varphi^2
+{32\over3\pi}|\varphi|^3\ln\Lambda
\right]+{2\over\pi}\varphi^2
\label{v1si}
\\
{\cal V}_1^{\pi}(\varphi) &=& 
{1\over 4\pi^2}\left\{
\int_0^{\Lambda}\!\! dy ~y^2\ln\left[{
|\varphi| - m_0 + 
\displaystyle{{y\over 2} {\rm tan}^{-1}
\left(\displaystyle{y \over 2 |\varphi| }\right)}
\over
\left|\pi y/4 - m_0\right|}
\right] - {16\over3\pi}|\varphi|^3\ln\Lambda
\right\}
\label{v1pi}
\end{eqnarray}
where $m_0=2-\pi/\lambda$ is the minimum
of ${\cal V}_0(\varphi)$ at $\alpha=0$. 
From this relation and from the study of the renormalization
group $\beta$-function it follows that the leading order
of the theory has a fixed point $\lambda^*=\pi/2$ (see Ref.\ \ \cite{ren})
which separates the two phases in which the chiral symmetry
is broken ($\lambda>\lambda^*$) and unbroken ($\lambda<\lambda^*$)
\cite{rosen0,rosen1,ren}. 
The order $1/N$ leads to a correction of the critical coupling of the form
\begin{equation}
{\bar\lambda}^{*}~=~\lambda^{*}~\left(1+{2\over N}\right)
\label{gcrit}
\end{equation}
valid for the zero temperature effective potential 
(\ref{zeroeffgen}) with only the scalar term (\ref{v1si}) and also when
the pseudoscalar (\ref{v1pi}) is included \cite{ren}. 

It is worth noticing that in studying the mean field approximation, 
it is not necessary to specify the value of $m_{0}$ or equivalently
of the coupling $\lambda$ (supposed to be $>\lambda^{*}$). 
It is in fact simple to verify that the effective potential can be
rescaled by $m_{0}$ by a re-definition of the fields and of the
mass parameter $\alpha$. This is no longer true 
at the order $1/N$. Thus, when studying the equation of state at order
$1/N$, although the gap equation needs to be solved
only at the mean field level, attention must be payed to choose a
value for the coupling $\lambda~>~{\bar\lambda}^{*}$,
namely consistent with chiral symmetry breaking at zero temperature
for $\alpha=0$. When presenting our results, it will be supposed
that $\lambda=\pi=2\lambda^{*}$ ~ (and thus $m_{0}=1$) which
is greater than ${\bar\lambda}^{*}$ for $N>2$. 

A potential difficulty of the expressions in Eqs.\ (\ref{v1si}) and
(\ref{v1pi}) is that they are not always real for small values
of $\varphi$ (see also \cite{rosen0}). Anyway they are real around
the mean field solution. This feature allows to evaluate the
pressure in a $1/N$ scheme. 

Notice that all the variables in the previous expressions are 
dimensionless. They have been rescaled by the (arbitrary) 
renormalization scale which serves as a subtraction point \cite{ren}. 


\section{Pressure at order ${\bf 1/N}$}
\label{sec:press}

According to the $1/N$ expansion derived in Ref.\ \cite{unosuenne}
the expression for the total pressure at order $1/N$ can be written as 
\begin{equation}
{\cal P}(T) = {\cal P}_0(\bar{\varphi}_0(T),T) + {1\over N}
\sum_{j=\sigma,\pi}{\cal P}_1^j(\bar{\varphi}_0(T),T)
\label{genpres}
\end{equation}
where ${\cal P}_0$ is the fermionic mean field term 
\begin{equation}
{\cal P}_0(\bar{\varphi}_0(T),T) = - \left[ {\cal V}_0(\bar{\varphi}_0(T))
- {\cal V}_0(\bar{\varphi}_0(0))\right] + {1\over\pi\beta}
\int_0^{+\infty}\!\!\!\!\!\!dx\ln\left(1+e^{\displaystyle 
-\beta\sqrt{x + \bar{\varphi}^2_0(T)} }\right)
\label{press0}
\end{equation}
The ${\cal P}_1^j$ terms, which contain the contribution of 
the bosonic fluctuations, can be divided for later convenience into two terms
\begin{equation}
{\cal P}_1^j\equiv{\cal P}^{j}_{1E} + {\cal P}^{j}_{1M}
\label{press1}
\end{equation}
where the subscripts $E$ and $M$ remind that the bosonic propagator
is evaluated for Euclidean and Minkowski four-momenta respectively. 
Their explicit forms are
\begin{eqnarray}
{\cal P}^{j}_{1E}\left(\bar{\varphi}_0,T\right)
&=& -\left[ {\cal V}_1^j(\bar{\varphi}_0(T))
-{\cal V}_1^j(\bar{\varphi}_0(0))\right]
\nonumber\\
&&-{1\over (2\pi)^2}\int_{0}^{+\infty}\!\!\!\!\!\!p~dp
\int_{0}^{+\infty}\!\!\!\!\!\!d\omega 
~\ln\left[1
+{i{\cal D}_{\beta,j}^{-1}(i\omega,{\bf p};\bar{\varphi}_0(T),T)
\over
i{\cal D}_{0,j}^{-1}(i\omega,{\bf p};\bar{\varphi}_0(T))} \right]
\label{press1e}
\end{eqnarray}
with ${\cal V}_1^j$ given in Eqs.\ (\ref{v1pi}), (\ref{v1si})
and
\begin{equation}
{\cal P}^{j}_{1M}=-\lim_{\epsilon\rightarrow0}
{1\over2\pi^2}\int_{0}^{+\infty}\!\!\!\!\!\!p~dp
\int_0^{+\infty}\!\!\!\!\!\!{d\omega}~
n_{B}(\omega)\left[{\rm arg}\left(
i {\cal D}_j^{-1}(\omega+i\epsilon,{\bf p}; \bar{\varphi}_0(T),T)\right)
-\pi\right]
\label{press1m}
\end{equation}
In the last expression, which is a pure
temperature correction, $n_{B}(\omega)$ is the Bose-Einstein 
distribution function
\begin{equation}
n_{B}(\omega)={1\over e^{\displaystyle \beta\omega} +1}
\end{equation}
and $arg(z)$ is the argument $\theta$ of the complex number 
$z=|z|\exp(i\theta)$. The expressions for the complete 
inverse bosonic propagators at finite temperature $i{\cal D}_{\beta ,j}^{-1}$
are given in Appendix\ \ref{sec:propag}. 

Notice that for a correct $1/N$ expansion of the pressure, both ${\cal P}_0$
and ${\cal P}_1$ are evaluated at the mean field solution $\bar{\varphi}_0(T)$
of the gap equation \cite{unosuenne,unosuenne1}, and
also that the bag constant
\begin{equation}
B\equiv {\cal V}_0(\bar{\varphi}_0(0),0)
+{1\over N}\sum_{j=\sigma,\pi}{\cal V}_1^j(\bar{\varphi}_0(0),0)
\end{equation}
has been subtracted from the beginning in order that
${\cal P}_0={\cal P}_1={\cal P}=0$ at $T=0$ (see Eqs.\ (\ref{press0}) 
and (\ref{press1e})). 

Let us first recall the results \cite{rosen,rosen1} 
which are obtained by extremizing the
finite temperature effective potential in the mean field
approximation, whose expression is given by minus the
pressure in Eq.\ (\ref{press0}) and with ${\bar \varphi}_{0} (T)$
taken, in this case, as a free variable. 

The gap equation, in $D=2+1$, can 
be solved analytically. One obtains

\begin{equation}
{\partial{\cal V}_{0}(\varphi,T)\over \partial\varphi}=
{\varphi\over \pi}~\left[|\varphi|~-~m_{0}~+~{2\over\beta}
\ln\left(1+e^{-\displaystyle{\beta |\varphi|}}\right)\right]
~+~\alpha~=~0
\label{meangap}
\end {equation}

Thus it is immediate to find that for $\alpha=0$, the
zero temperature solution outside the origin 
${\bar\varphi}_{0}(0) = m_{0}$ merges continuously into the
one located in the origin at the second order critical temperature
\begin{equation}
T_{c} = {m_{0}\over 2 \ln 2}\simeq0.7213~m_0
\label{ticci}
\end{equation}

Of course, the $1/N$ correction spoils the validity of
chiral symmetry breaking in the case of a continuous symmetry
(\ref{contsim}). As already said, we have considered the latter
only for $\alpha\neq 0$. 

In Eq.\ (\ref{press1e}) we have separated the zero temperature effective 
potential (which depends on temperature only through $\bar{\varphi}_0(T)$)
from the second term, which depends explicitly on temperature. 

Apparently ${\cal P}^{j}_{1E}$ become ill defined above a certain 
value of the temperature, since ${\bar\varphi}_{0}(T)$ decreases
for increasing temperatures and, as already said, both ${\cal V}_1^{\sigma}$
and ${\cal V}_1^{\pi}$	in (\ref{v1si}) and (\ref{v1pi}) exhibit an
imaginary part for small arguments. Furthermore, also the finite
temperature corrections to them (the logarithms in Eq.\ (\ref{press1e}))
turn out to be complex. 
Nevertheless we have verified that their sums,
which can be written as follows
\begin{equation}
{\cal P}^{j}_{1E}=
-{1\over (2\pi)^2}\int_{0}^{+\infty}\!\!\!\!\!\!p~dp
\int_{0}^{+\infty}\!\!\!\!\!\!d\omega 
~\ln\left[
i{\cal D}_{0,j}^{-1}(i\omega,{\bf p};\bar{\varphi}_0(T))
+ i{\cal D}_{\beta,j}^{-1}(i\omega,{\bf p};\bar{\varphi}_0(T),T)
\right] + c. t. 
\end{equation}
are always real when evaluated at the minimum $\bar{\varphi}_0(T)$
of the mean field effective potential. This is a very important 
feature since it ensures that the total pressure is a well defined 
quantity even at order $1/N$. 

Finally, the ${\cal P}^{j}_{1M}$ terms in Eq.\ (\ref{press1m}) are
pure temperature terms, whose interpretation is more direct, at least
in case of a free propagation, where they correspond to the
pressure of a gas of free bosons. 

\section{Numerical results}
\label{sec:results}

\subsection{Discrete Symmetry}

By handling the expressions for $i{\cal D}^{-1}_{\beta ,j}$ 
as shown in Appendix\ \ref{sec:propag}, 
the number of successive integrations necessary
to calculate the pressure in Eq.\ (\ref{genpres})
is reduced, and thus the study can be better completed numerically
by means of suitable routines of integration. 
The preliminary treatment of $i{\cal D}^{-1}_{\beta ,j}$ 
is important expecially to calculate ${\cal P}^{j}_{1E}$, 
which are slowly convergent expressions for large momenta. 

Figs.\ \ref{fig1}-\ref{fig3} summarize the behaviour of the total
pressure for the discrete symmetry model ($\lambda_{5}=0$)
in the chiral limit $M=0$ ($\alpha=0$). 
With only the scalar composite included, the 
total pressure in Eq.\ (\ref{genpres}) is composed of 
three terms
\begin{equation}
{\cal P}~ =~{\cal P}_{0}+{1\over N}{\cal P}^{\sigma}_{1}
~= ~{\cal P}_{0} + {1\over N}\left({\cal P}^{\sigma}_{1E}
+{\cal P}^{\sigma}_{1M}\right) 
\label{genpressi}
\end{equation}
The results are given for $\lambda=\pi$,
thus $m_{0}=1$ and from Eq.\ (\ref{ticci})
the mean field critical temperature is $T_{c}=(2~\ln2)^{-1}$. 
Since $T_{c}$ sets the natural scale of low and high temperatures, 
all the temperature behaviours are plotted vs. $T/T_{c}$. 
As usual, in Figs.\ \ref{fig1}-\ref{fig3} the pressure is divided 
by $T^{D}=T^{3}$ in order to make evident the Stefan-Boltzmann regime 
at high temperatures. 
In Fig.\ \ref{fig1} the mean field term ${\cal P}_{0}$
(stars) is compared with
${\cal P}^{\sigma}_{1E}$ (empty circles) and
${\cal P}^{\sigma}_{1M}$ (full circles) separately. 
Notice that while ${\cal P}^{\sigma}_{1M}$ remains smaller 
than the mean field term in the whole range
of temperatures, ${\cal P}^{\sigma}_{1E}$ is large for
intermediate temperatures (even if it has still to be divided by $N$),
and it is slightly negative for $T{\ \lower -2.5pt\hbox{$>$} 
\hskip-8pt \lower 2.5pt \hbox{$\sim$}\ } 1.7~T_{c}$. 
In Fig.\ \ref{fig2} the mean field term is compared to
the total contribution of order $1/N$ for $N=1$, 
${\cal P}^{\sigma}_{1}$, which comes out to be positive
in the whole range of temperatures. 

In Fig.\ \ref{fig3} we plot the total pressure of 
Eq.\  (\ref{genpressi}) divided by $T^3$, for $N=3$ 
(full triangles), $N=10$ (full circles)
and $N=\infty$ (stars).
The curves show that in absence of pion-like
excitations, the low-temperature behaviour is not
much affected by the $1/N$ correction, as it was expected,
whereas in the region of intermediate temperatures
(around $T_{c}$) this correction is still large with
$N=10$, suggesting the importance of higher orders in $1/N$.
Finally, we notice that the contribution of the
scalar composite starts to be negligible at about twice the
mean field critical temperature.

\subsection{Continuous Symmetry}

In Figs.\ \ref{fig4}-\ref{fig6} we plot the behaviour of the
pressure divided by $T^3$ (see Eqs.\ (\ref{genpres}) and (\ref{press1})) 
for the continuous symmetry model ($\lambda_{5}=\lambda$),
for nonvanishing fermion masses ($\alpha=0.001$ in 
Figs.\ \ref{fig4}-\ref{fig6}(a) and $\alpha=0.01$ in 
Figs.\ \ref{fig4}-\ref{fig6}(b)). 
All the curves are plotted vs. $T/T_{c}$ as done in the discrete
symmetry case ($T_{c}=(2\ln2)^{-1}$).

In Fig.\ \ref{fig4} we compare the mean field term ${\cal P}_{0}$
(stars) with
${\cal P}^{\sigma}_{1E}$ (empty circles),
${\cal P}^{\sigma}_{1M}$ (full circles), 
${\cal P}^{\pi}_{1E}$ (empty triangles) and
${\cal P}^{\pi}_{1M}$ (full triangles). 
This figure show that, for very low temperatures and small values of $\alpha$, 
the pressure is dominated by the pion contribution. We will further comment
on this point when referring to Fig.\ \ref{lowpisi}.
Notice also that ${\cal P}^{\pi}_{1M}$ and ${\cal P}^{\sigma}_{1M}$ 
become almost degenerate for $T > T_{c}$. In fact the inverse propagators
$i{\cal D}^{-1}_{\sigma}$ and $i{\cal D}^{-1}_{\pi}$ in Eq.\ (\ref{press1m})
are equal for $\bar{\varphi}_0(T)=0$, and even if, after inserting
a bare fermion mass, this condition is never realized, for small $\alpha$,
$\bar{\varphi}_0(T)\sim \alpha$, and therefore
 the difference between ${\cal P}^{\pi}_{1M}$ and ${\cal P}^{\sigma}_{1M}$
 is negligible on the scale of the figures.

In Fig.\ \ref{fig5} the mean field term is compared to
the total contribution of ${\cal P}^{\sigma}_{1}$ and ${\cal P}^{\pi}_{1}$.

In Fig.\ \ref{fig6} we plot of the total pressure in Eq.\ (\protect\ref{genpres})
for $N=3$ (full triangles), $N=10$ (full circles) and $N=\infty$ (stars). 

We remark that, although ${\cal P}^{\pi}_{1}$ is negative in a finite
range of temperatures below $T_{c}$, the total pressure of order $1/N$
is always positive, and that its contribution is still large in the 
critical region, even for $N=10$. As in the discrete symmetry version of the model
the role of composites becomes negligible at $T\simeq 2T_c$

As we have just pointed out, Figs.\ \ref{fig4}-\ref{fig6}
show that, for very low temperatures, the pressure is dominated by
the pion contribution (at least for small values of $\alpha$). 
This is in agreement with what one expects on general grounds 
since the contribution of massive
particles to the partition function is exponentially depressed 
as $\exp(-m/T)$ for $T\rightarrow0$, and therefore
the pion, being the lightest particle in the spectrum, 
dominates the other terms \cite{leutw}. 

In Fig.\ \ref{fig4} it is also evident that
the dominant pion contribution comes in particular from the term
${\cal P}_{1M}^{\pi}$ given in Eq.\ (\ref{press1m}). 
It is easy to verify that this term reduces to the standard 
expression for the pressure of a free gas, in case of 
$i{\cal D}^{-1}=\omega^2 -{\bf p}^2 -m^2$. 
Thus it is interesting to compare our results with the free
gas approximation in order to quantify, in the low temperatures region,
the contribution of the pion and sigma poles to the total pressure. 
In the case of free particles the expression for the pressure density is
\begin{eqnarray}
{{\cal P}\over V} &=& \pm g T \int{d^2 p\over (2\pi)^2}
\ln\left(1\pm{\rm e}^{\displaystyle -\sqrt{{\bf p}^2 + m^2}/T}\right)
\nonumber\\
&=&{g T^3\over 2\pi}\sum_{n=1}^{+\infty}(\mp1)^{n+1}\frac{1}{n^2}
\left(\frac{1}{n}+\frac{m}{T}\right){\rm e}^{\displaystyle -n m/T}
\end{eqnarray}
where $g$ is a degeneration factor and $\pm$ refer to fermions and bosons 
respectively. 

In Fig.\ \ref{lowpisi} we compare, for various values of $\alpha$,
the low temperature behaviour of ${\cal P}_0$, ${\cal P}_{1M}^{\sigma}$
and ${\cal P}_{1M}^{\pi}$
(divided by $T^3$) with that of a free particles 
gas of mass $m=|\bar{\varphi}_0(0)|$, $m_{\sigma}$ and $m_{\pi}$ respectively
(all the masses are evaluated at $T=0$). 
For the fermionic (mean field) term we have 
$g=4$ (see Eq.\ (\ref{press0})) and $m=|\bar{\varphi}_0(0)|$, whereas 
for the bosons the degeneration factor is $g=1$ and
$m_{\sigma}$ and $m_{\pi}$ are defined through the zeros of the 
inverse propagator in Eq.\ (\ref{dzero}). 
The curves are plotted for $\alpha=0, 0.001, 0.01$. 
The pion pressure for $\alpha=0$ is plotted as a reference level, even though
in this model (with continuous symmetry) the zero mass limit is not allowed 
since the pion cannot exist as
Goldstone particle in $d=2$ spatial dimensions at finite temperature. 
Notice that,
whereas the pion term changes dramatically when $\alpha$ is turned on,
the fermionic and sigma terms for $\alpha=0.001$
and $\alpha=0.01$ are, on this scale, almost undistinguishable from the 
case $\alpha=0$. 

From Fig.\ \ref{lowpisi} it is evident that, for very low temperatures
($T {\ \lower -2.5pt\hbox{$<$} \hskip-8pt \lower 2.5pt \hbox{$\sim$}\ }
0.2~T_c$), the data of our model are very well fitted by the 
free gas approximation. 
This implies that in this region, although the boson propagator
has a complex structure (see Appendix\ \ref{sec:propag}), the pole gives 
almost the whole contribution. For the fermions the approximation
works even better due to the fact that the only deviation from a free gas
behaviour relies on the temperature dependence of the condensate 
$\bar{\varphi}_0$, which is almost constant for low $T$. 

Notice also that, when the exact curves of ${\cal P}_{1M}^{j}$ 
start deviating from the free gas 
behaviour, the contribution of ${\cal P}_{1E}^{\pi}$ and 
${\cal P}_{1E}^{\sigma}$
is already important (see Figs.\ \ref{fig4}(a)-(b)) and
the approximation is no longer reliable. In this model this
happens for temperatures of the order of $0.1~T_c$.

\section{Conclusions}
\label{sec:concl}
The paradigm of chiral restoration beyond a critical temperature and the 
ensuing physical implications is a crucial aspect of chiral theories, engendering
at this time intensive studies. The most relevant physical case
in high energy is QCD, but its complexity makes its complete
study still very difficult. For certain aspects, expecially
concerning chirality, the Gross-Neveu model is believed
to have features similar to QCD.

We have here performed a calculation of the thermodynamics of the
Gross-Neveu model in 2+1 dimensions (both in its discrete symmetry
and continuous symmetry versions), at order $1/N$, for the whole
range of temperatures below and after the chiral phase transition.
The calculation is performed by specializing a general formula
valid for four-fermion theories. Renormalization is treated
consistently.

The possibility of a detailed calculation for this model allows
for quantitative comparisons of the different contributions to
the equation of state in the various temperature regions, below, near,
and after the transition. In particular we find that the intuitive
expectations on dominance of the pion-like excitations at low $T$,
and on free-component behaviour at large $T$ are satisfied. On the
other hand we are able to quantitatively assess the limits for the
validity of such behaviours. Pion-like dominance holds only for
$T$ much smaller than the critical temperature, and free-component
behaviour only from at least twice the critical temperature.
Such limitations hold even for sufficiently large $N$. Within
the intermediate region around the phase transition temperature,
the role of composite states appears important, and near the
critical temperature it might be unavoidable to go to higher
$1/N$ orders. We also find quantitative differences between the
cases of vanishing and non-vanishing fermion masses.

\acknowledgements{We thank A. Barducci and R. Casalbuoni for interesting
discussion. 
This work has been carried out within the
EEC program Human Capital and Mobility (N. OFES 95.0200 and CHRXCT 94-0579). 
}

\appendix
\section{Bosonic propagator}
\label{sec:propag}

In this Appendix we discuss the structure of the inverse bosonic
propagator, which can be written as the sum of the zero temperature
term, $i{\cal D}_0^{-1}$, and a temperature dependent part, 
$i{\cal D}_{\beta}^{-1}$ (for convenience we omit the index $j$)
\begin{equation}
i{\cal D}^{-1}(\omega, {\bf p}; \varphi, T) \equiv 
i{\cal D}_0^{-1}(\omega, {\bf p}; \varphi) + 
i{\cal D}_{\beta}^{-1}(\omega, {\bf p}; \varphi, T)
\end{equation}

The zero temperature inverse bosonic propagator $i{\cal D}_0^{-1}$
is \cite{rosen,unosuenne}
\begin{eqnarray}
i{\cal D}^{-1}_{0}(\omega, {\bf p}; \varphi)
&=& - (1+\delta Z_2) + 4i\lambda
\int{d^3 q\over(2\pi)^3}\displaystyle{q_{\mu}
(q^{\mu}+p^{\mu})\pm \varphi^2\over
\left[(q_{\mu}+p_{\mu})(q^{\mu}+p^{\mu})-\varphi^2\right]
(q_{\mu}q^{\mu} - \varphi^2)}
\end{eqnarray}
where $\pm$ refer to the scalar and pseudoscalar fields respectively,
$p_{\mu}=(\omega,{\bf p})$ and $\delta Z_2$ is the counterterm 
needed for the renormalization of the one loop effective potential
(see Sect.\ \ref{sec:effpot}). 
The renormalized expression is
\begin{eqnarray}
i{\cal D}_{0}^{-1}\left(\omega, {\bf p}; \varphi\right)&=&
{\lambda\over\pi}
\theta({\bf p}^2 -\omega^2)\Bigg\{
m_0 - |\varphi| + {1\over2} {\omega^2 - {\bf p}^2 - \epsilon^2_M\over
\sqrt{{\bf p}^2 -\omega^2}}\tan^{-1}
\sqrt{{\bf p}^2 -\omega^2\over 4\varphi^2}\Bigg\}
\label{dzero}
\\
&&+{\lambda\over\pi}\theta(\omega^2 - {\bf p}^2)
\left\{m_0 - |\varphi| + {1\over4} 
{\omega^2 - {\bf p}^2-\epsilon^2_M\over\sqrt{\omega^2 - {\bf p}^2}}
\ln\left|{ 2|\varphi| + \sqrt{\omega^2 - {\bf p}^2} 
\over 2|\varphi| - \sqrt{\omega^2 - {\bf p}^2}}\right| \right\}
\nonumber\\
&&
-i{\lambda\over\pi}\theta\left(\omega^2 - {\bf p}^2 - 4\varphi^2\right)
 {\pi\over4} ~
{\omega^2 - {\bf p}^2-\epsilon^2_M\over\sqrt{\omega^2 - {\bf p}^2}}
\nonumber
\end{eqnarray}
where $\epsilon_{M_{\sigma}}^2 = 4 \varphi^2$ and 
$\epsilon_{M_{\pi}}^2 = 0$. 

At finite temperature, the inverse propagator given in 
Ref.\ \cite{unosuenne}, can be cast in a more compact form as
\begin{eqnarray}
i{\cal D}_{\beta}^{-1}(\omega, {\bf p}; \varphi, T) &=& 
4\lambda\int{d^2 q\over (2\pi)^2}
{\displaystyle n_{F}(E_q)\over E_q}
+ \lambda\left(\omega^2 -{\bf p}^2 - \epsilon^2_M\right)
\cdot\nonumber\\
&&\qquad\cdot
\int{d^2 q\over (2\pi)^2}
{\displaystyle n_{F}(E_q)\over E_q}
\sum_{\xi_1,\xi_2=\pm1}
{\displaystyle 1\over 
(E_q + \xi_1 \omega)^2 - E_{q+\xi_2 p}^2}
\label{dbeta}
\end{eqnarray}
where $E_q\equiv\sqrt{{\bf q}^2 + \varphi^2}$, and $n_{F}(E_q)$ is the
Fermi-Dirac distribution function
\begin{equation}
n_{F}(E_q) = {1\over {\rm e}^{\displaystyle\beta E_q} + 1}
\end{equation}
To calculate the real and imaginary part of Eq.\ (\ref{dbeta}) 
we evaluate 
$i{\cal D}_{\beta}^{-1}(\omega + i\epsilon, {\bf p}; \varphi, T)$
in the limit $\epsilon\rightarrow0$. Therefore, by using the formula
\begin{eqnarray}
\lefteqn{
\lim_{\epsilon\rightarrow 0}
\int_{0}^{2\pi}d\theta{1\over A + B \cos\theta + i \alpha\epsilon}
=}&&
\nonumber\\
&&\qquad\qquad{2\pi\over\sqrt{|A^2 - B^2|}}
\left[{\rm sign}(A)\theta\left(A^2 - B^2\right)
-i{\rm sign}(\alpha)\theta\left(B^2 - A^2\right)\right]
\end{eqnarray}
we get ($x \equiv E_q$)
\begin{eqnarray}
 i{\cal D}_{\beta}^{-1}\left(\omega, {\bf p}; \varphi, T\right)
&=&{\lambda\over\pi}\Bigg\{-{2\over\beta}
\ln\left(1+{\rm e}^{\displaystyle-\beta|\varphi|}\right)
+\left(\omega^2 - {\bf p}^2 -\epsilon^2_M\right)
\sum_{\xi_1=\pm1}\int_{|\varphi|}^{+\infty}
dx~\cdot\nonumber\\
&&\cdot{n_F(x)\over\sqrt{|A^2 - B^2|}}
\left[{\rm sign}(A)\theta\left(A^2 - B^2\right)
-i{\rm sign}(\omega+\xi_1 x)\theta\left(B^2 - A^2\right)\right]\Bigg\}
\end{eqnarray}
with
\begin{eqnarray}
A&=&(\omega^2 - {\bf p}^2)+2\xi_1\omega x 
\\
A^2 - B^2 &=& 4(\omega^2 - {\bf p}^2)x^2 + 4\xi_1\omega
(\omega^2 - {\bf p}^2)x +
(\omega^2 - {\bf p}^2)^2 + 4 {\bf p}^2\varphi^2
\label{a2b2}
\end{eqnarray}
By studying the sign of $A^2 - B^2$ in Eq.\ (\ref{a2b2}), 
the $\theta$-functions can be implemented in
the integration extrema as follows
\begin{eqnarray}
&&\int_{|\varphi|}^{+\infty}\!\!dx
\theta\left(A^2 - B^2\right) = \theta\left({\bf p}^2 - \omega^2\right)
\int_{|\varphi|}^{p\sqrt{\Delta} - \xi_1\omega\over2}\!\!dx
+ \theta\left( (\omega^2 - {\bf p}^2)
({\bf p}^2 - \omega^2 + 4\varphi^2)\right)
\int_{|\varphi|}^{+\infty}\!\!dx
\nonumber\\
&&\qquad
+ \theta\left( \omega^2 - {\bf p}^2 - 4\varphi^2 \right)
\left[\
\delta_{\xi_1,1} \int_{|\varphi|}^{+\infty}\!\!dx
+ \delta_{\xi_1,-1}\left(
\int_{|\varphi|}^{\omega - p\sqrt{\Delta}\over2 }\!\!dx+
\int_{\omega + p\sqrt{\Delta}\over2 }^{+\infty}\!\!dx
\right)
\right]
\\
\nonumber\\
&&\int_{|\varphi|}^{+\infty}\!\!dx
\theta\left(B^2 - A^2\right) = \theta\left({\bf p}^2 - \omega^2\right)
\int_{p\sqrt{\Delta} - \xi_1\omega\over2}^{+\infty}\!\!dx
+ \theta\left( \omega^2 - {\bf p}^2 - 4\varphi^2 \right)
\delta_{\xi_1,-1}\int_{\omega - p\sqrt{\Delta}\over2 }^{
\omega + p\sqrt{\Delta}\over2 }\!\!dx
\end{eqnarray}
where we have defined $\Delta \equiv 1 -4\varphi^2/
\left( \omega^2 - {\bf p}^2\right)$

The expression of the inverse propagator evaluated for
imaginary values of the energy, $\omega\rightarrow i\omega$,
needed in Eqs.\ (\ref{poteff_nr}) and (\ref{press1m}), 
follows from Eqs.\ (\ref{dzero}) and (\ref{dbeta}). 
At zero temperature, from Eq.\ (\ref{dzero}), we easily get
\begin{equation}
 i{\cal D}_{0}^{-1}\left(i\omega, {\bf p}; \varphi\right) =
{\lambda\over\pi}\Bigg\{
m_0 - \varphi - {1\over2} {\omega^2 + {\bf p}^2+\epsilon^2_M
\over\sqrt{\omega^2 + {\bf p}^2}}\tan^{-1}
\sqrt{\omega^2 + {\bf p}^2\over 4\varphi^2}\Bigg\}
\end{equation}
whereas at finite temperature, by performing the angular integration in 
Eq.\ (\ref{dbeta}), we have
\begin{equation}
 i{\cal D}_{\beta}^{-1}\left(i\omega, {\bf p}; \varphi, T\right)
={\lambda\over\pi}
\int_{|\varphi|}^{+\infty} dx~n_F(x)
\left[\sqrt{2}\left(\omega^2 + {\bf p}^2 + \epsilon^2_M\right)
{\sqrt{\rho+a^2 - b^2 - c^2 }\over\rho} - 2\right]
\end{equation}
 with
\begin{eqnarray}
a^2 \equiv (\omega^2 + {\bf p}^2)^2 \; ;\qquad
b^2 &\equiv& 4 {\bf p}^2 {\bf q}^2 = 
4 {\bf p}^2 x^2 - 4 {\bf p}^2 \varphi^2 \; ;\qquad 
c^2 \equiv 4 w^2 x^2 \\
\rho &\equiv& \sqrt{(a^2 - b^2 - c^2)^2 + 4 a^2 c^2}
\end{eqnarray}

\begin{figure}
\caption{Plot of the pressure
entering Eq.\ (\protect\ref{genpressi}), divided by $T^3$,
vs. $T/T_{c}$ ($T_{c}=(2\ln2)^{-1}$)
for the discrete symmetry model ($\lambda_5=0$)
with zero current fermion masses. 
The mean field term ${\cal P}^{\sigma}_{0}/T^3$ (stars) is
compared to the $1/N$ correction
for $N=1$, which is divided into two terms:
${\cal P}^{\sigma}_{1E}/T^3$(open circles) and ${\cal P}^{\sigma}_{1M}/T^3$
(full circles). The low temperature
behaviours are enhanced in the right upper part of the figure. 
\label{fig1}}
\end{figure}

\begin{figure}
\caption{Plot of the pressure divided by $T^3$
vs. $T/T_{c}$ in Eq.\ (\protect\ref{genpressi}) ($T_{c}=(2\ln2)^{-1}$),
 for the discrete symmetry model with zero current fermion masses. 
The mean field ${\cal P}^{\sigma}_{0}/T^3$ (stars) is compared
to the $1/N$ correction for $N=1$, ${\cal P}^{\sigma}_{1}/T^3$ 
(full circles). The low temperature
behaviours are enhanced in the right upper part of the figure. 
\label{fig2}}
\end{figure}

\begin{figure}
\caption{Plot of the total pressure divided by $T^3$
vs. $T/T_{c}$ in Eq.\ (\protect\ref{genpressi}) ($T_{c}=(2\ln2)^{-1}$), for
the discrete symmetry model with $\alpha=0$. 
The curves represent ${\cal P}/T^3$ for $N=3$ (full triangles),
$N=10$ (full circles) and $N=\infty$ (stars). 
The low temperature
behaviours are enhanced in the right lower part of the figure. 
\label{fig3}}
\end{figure}

\begin{figure}
\caption{Plots of the pressure divided by $T^3$
vs. $T/T_{c}$ in Eqs.\ (\protect\ref{genpres}), (\protect\ref{press1}) 
($T_{c}=(2\ln2)^{-1}$), for
the continuous symmetry model with a bare fermion mass term. 
Figs.\ (a,b) are for $\alpha=0.001$ and $\alpha=0.01$ respectively. 
The curves represent ${\cal P}^{\sigma}_{1E}/T^3$ (open circles),
${\cal P}^{\sigma}_{1M}/T^3$ (full circles),
${\cal P}_{0}/T^3$ (stars), ${\cal P}^{\pi}_{1M}/T^3$ (full triangles)
and ${\cal P}^{\pi}_{1E}/T^3$ (open triangles). 
${\cal P}^{\pi}_{1M}/T^3$ and ${\cal P}^{\sigma}_{1M}/T^3$ 
become (almost) degenerate after $T_{c}$, the difference being 
undistinguishable on this scale. 
The low temperature
behaviours are enhanced in the right upper part of the figure. 
\label{fig4}}
\end{figure}

\begin{figure}
\caption{Plots of the pressure divided by $T^3$
vs. $T/T_{c}$ in Eq.\ (\protect\ref{genpres}) ($T_{c}=(2\ln2)^{-1}$), for
the continuous symmetry model with a bare fermion mass term. 
Figs.\ (a,b) are for $\alpha=0.001$ and $\alpha=0.01$ respectively. 
The curves compare the mean field approximation
${\cal P}_{0}/T^3$ (stars) with the total $1/N$ corrections
for $N=1$ of the scalar composite ${\cal P}^{\sigma}_{1}/T^3$ (full circles),
and of the pseudoscalar ${\cal P}^{\pi}_{1}/T^3$ (full triangles)
The low temperature
behaviours are enhanced in the right upper part of the figure. 
\label{fig5}}
\end{figure}

\begin{figure}
\caption{Plot of the total pressure divided by $T^3$
vs. $T/T_{c}$ in Eq.\ (\protect\ref{genpres}) ($T_{c}=(2\ln2)^{-1}$), for
the continuous symmetry model with a bare fermion mass term. 
Figs.\ (a,b) are for $\alpha=0.001$ and $\alpha=0.01$ respectively. 
The curves represent ${\cal P}/T^3$ for $N=3$ (full triangles),
$N=10$ (full circles) and $N=\infty$ (stars). 
The low temperature
behaviours are enhanced in the right lower part of the figure. 
\label{fig6}}
\end{figure}

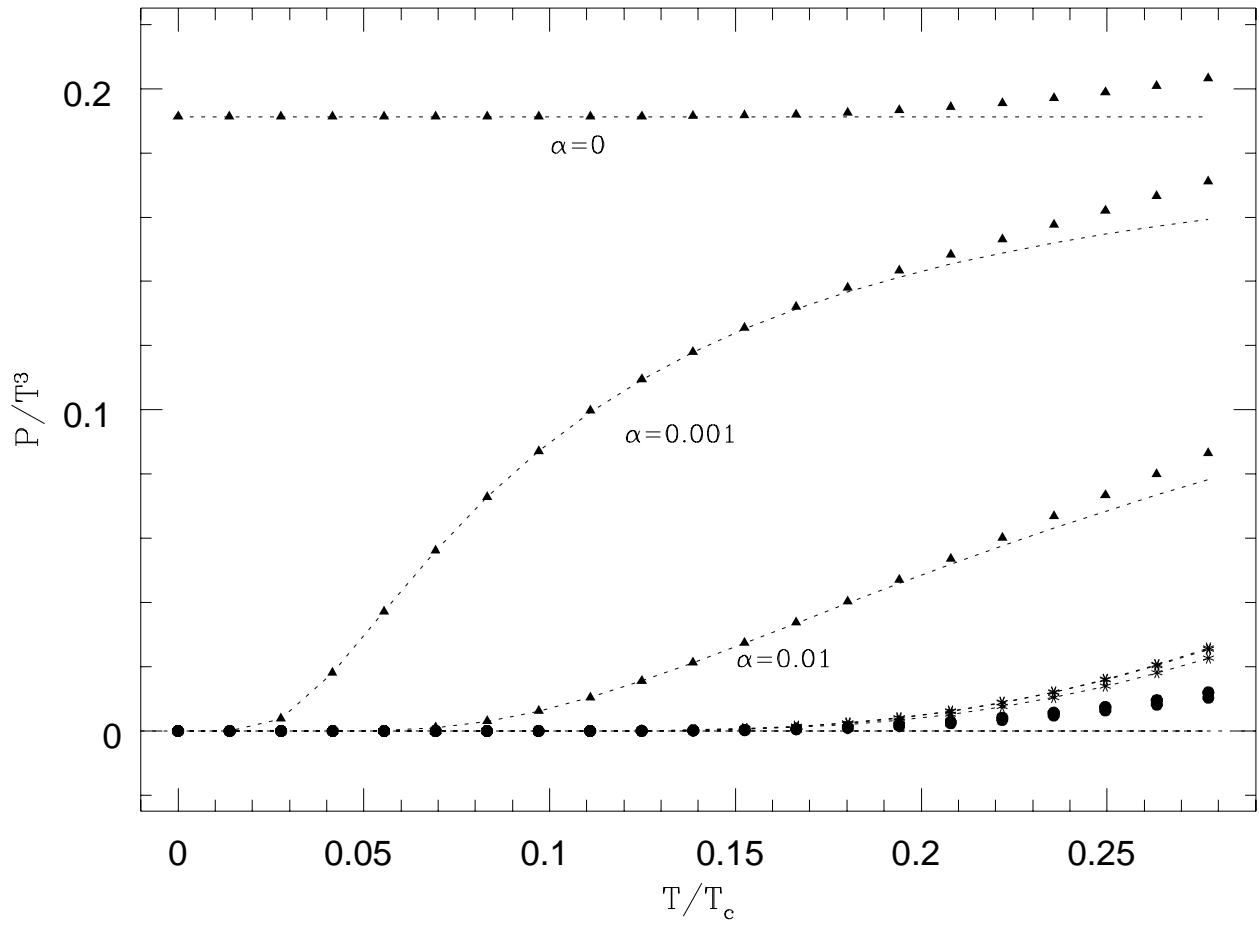
\begin{figure}
\caption{Comparison between the low temperature behaviour of 
${\cal P}_0/T^3$ (stars), ${\cal P}_{1M}^{\sigma}/T^3$ 
(full circles)
and ${\cal P}_{1M}^{\pi}/T^3$ (full triangles) 
and that of a free particles gas
of masses $m=|\bar{\varphi}_0(0)|$, $m_{\sigma}$ and $m_{\pi}$ respectively 
(dotted lines)
for $\alpha=0, 0.001, 0.01$ ($T_c=(2\ln2)^{-1}$). 
The pion pressure for $\alpha=0$ is plotted as a reference level, even though
in this model (with continuous symmetry) the zero mass limit is not allowed. 
Notice that,
whereas the pion term changes dramatically when $\alpha$ is turned on,
the fermionic and sigma terms for $\alpha=0.001$
and $\alpha=0.01$ are, on this scale, almost undistinguishable from the 
case $\alpha=0$. 
\label{lowpisi}}
\end{figure}

\end{document}